\documentclass[lettersize,journal]{IEEEtran}

\usepackage[utf8]{inputenc}
\usepackage[T1]{fontenc}
\usepackage{amsmath,amssymb,amsfonts}
\usepackage{algorithm}
\usepackage{algorithmic}
\usepackage{array}
\usepackage{graphicx}
\usepackage{cite}
\usepackage{booktabs}
\usepackage{multirow}
\usepackage{xcolor}
\usepackage{tikz}
\usetikzlibrary{shapes,arrows,positioning,calc}
\usepackage{tabularx}
\usepackage{booktabs}
\usepackage{array}

\newcolumntype{Y}{>{\raggedright\arraybackslash}X}

\usepackage{hyperref}
\hypersetup{
    hidelinks
}

\renewcommand{\arraystretch}{1.20}
\footnotesize

\begin{document}

\title{LLM-Based Multi-Agent Systems over Wireless Networks: A Joint Agent--Network Design Perspective}

\author{Chao Hu,~\IEEEmembership{Graduate Student Member,~IEEE,}
Yuan Guo,~\IEEEmembership{Member,~IEEE,}
Guanlin Wu,~\IEEEmembership{Graduate Student Member,~IEEE,}
Yueling Che,~\IEEEmembership{Member,~IEEE,}
Han Hu,~\IEEEmembership{Senior Member,~IEEE,}
and Jie Xu,~\IEEEmembership{Fellow,~IEEE}

\thanks{C. Hu is with the College of Computer Science and Software Engineering (CSSE), Shenzhen University, Shenzhen 518060, China, and also with the School of Science and Engineering (SSE), the Shenzhen Future Network of Intelligence Institute (FNii), and the Guangdong Provincial Key Laboratory of Future Networks of Intelligence, The Chinese University of Hong Kong (Shenzhen), Shenzhen 518172, China (e-mail: chaohu@link.cuhk.edu.cn).}

\thanks{Y. Guo, G. Wu, and J. Xu are with the SSE, the FNii, and the Guangdong Provincial Key Laboratory of Future Networks of Intelligence, The Chinese University of Hong Kong (Shenzhen), Shenzhen 518172, China (e-mail: guoyuan@cuhk.edu.cn; guanlinwu1@link.cuhk.edu.cn; xujie@cuhk.edu.cn).}

\thanks{Y. Che is with the CSSE, Shenzhen University, Shenzhen 518060, China (e-mail: yuelingche@szu.edu.cn).}

\thanks{H. Hu is with the School of Information and Electronics, Beijing Institute of Technology, Beijing 100081, China (e-mail: hhu@bit.edu.cn).}

\thanks{Y. Guo and Y. Che are the corresponding authors.}}
\maketitle

\begin{abstract}
As large language models (LLMs) evolve from standalone models into collaborative agents embedded in physical systems, their reasoning and execution are increasingly distributed across wireless edge nodes. In this setting, wireless networks are experiencing a paradigm shift from only providing data connectivity to supporting the multi-agent reasoning workflow itself. The task performance of such network-constrained LLM-based multi-agent systems (MASs) is jointly affected by the multi-agent reasoning dependencies as well as the underlying network connectivity and edge resources. This coupling gives rise to various technical challenges, including the metric misalignment and message redundancy, state inconsistency and topology mismatch, as well as resource limitation and trust discontinuity. To address these challenges, this article develops a novel joint agent--network design perspective that coordinates decisions on both sides of the system. Specifically, we present the joint design of agent--interaction scheduling and resource allocation, the message selection-transmission co-design, as well as the joint agent--network topology design and workload--resource allocation. Furthermore, we consider the network-verified provenance that is linked with agent-side information-flow control to constrain how received information affects subsequent operations. An illustrative vehicle-to-everything (V2X) case study shows that jointly adapting agent-side interaction decisions and network operations improves task completion under communication and edge-resource constraints, outperforming the conventional agent-only and wireless-only separate designs.
\end{abstract}


\begin{figure*}[t]
\centering
\includegraphics[width=0.99\textwidth]{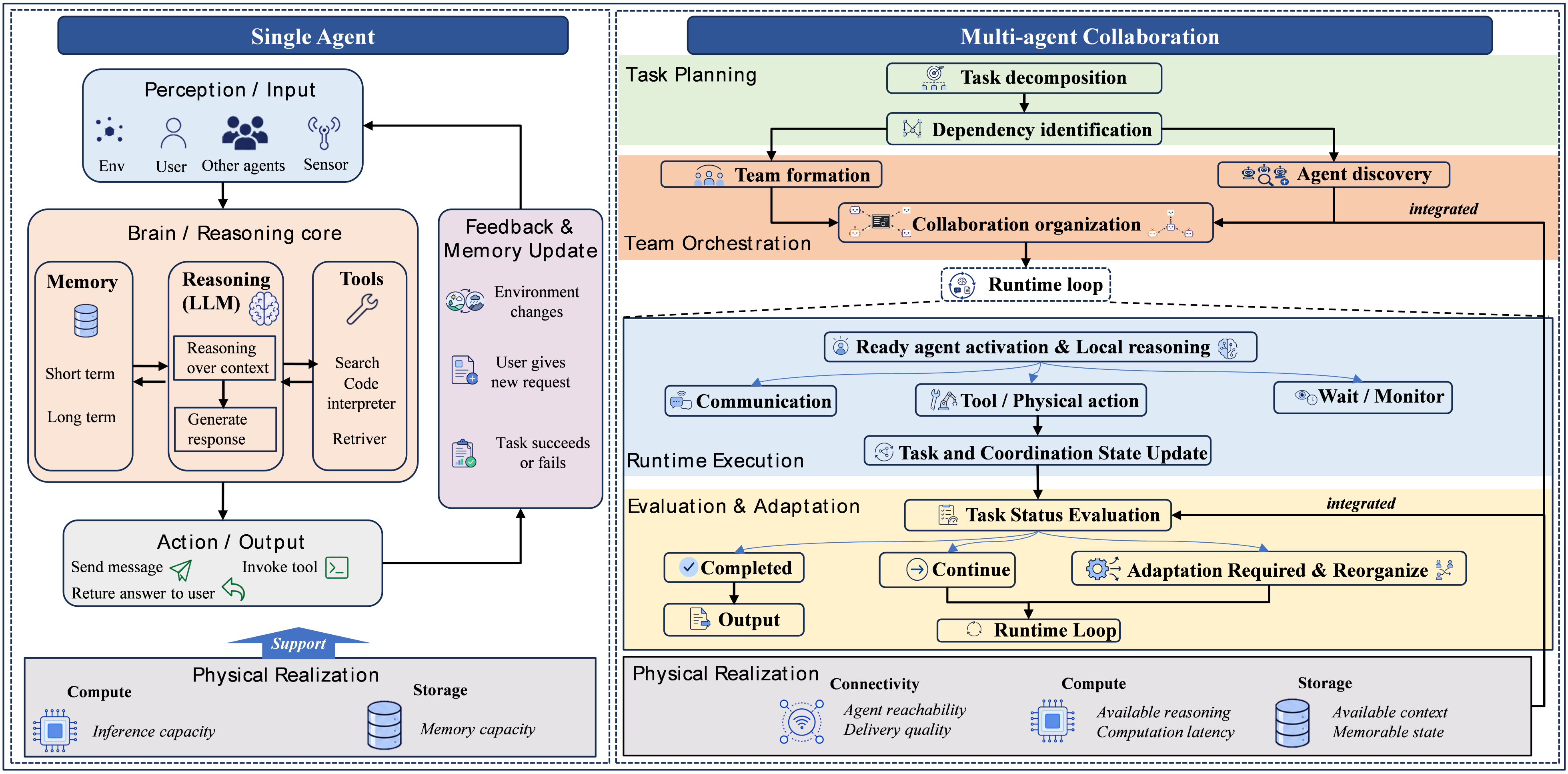}
\caption{From single-agent operation to multi-agent collaboration over wireless networks. Task planning forms inter-agent collaboration, while its runtime execution and adaptation depend on wireless connectivity and edge resources.}
\label{fig:system}
\end{figure*}

\section{Introduction}

Large language models (LLMs) are extending artificial intelligence (AI) from passive inference toward autonomous agents. These agents can reason over context, invoke external tools, and act toward specific objectives~\cite{llm_agent_survey,autogen}. As tasks become more complex with heterogeneous information and reasoning
requirements, a single agent may become insufficient. This limitation motivates LLM-based multi-agent systems (MASs), where the reasoning process is distributed across multiple agents with complementary roles~\cite{jiang2024,darwin_godel}.

On the other hand, these MASs are increasingly deployed in distributed physical systems, where agents rely on observations collected at different locations and must update their decisions under time-critical requirements. In this case, conventional cloud-centric architectures, in which the reasoning needs to be implemented at far-apart cloud nodes with increased communication and execution delay, are poorly matched to multi-agent reasoning. By contrast, edge computing is becoming particularly appealing, which places reasoning at edge nodes close to the observation sources, allowing newly acquired observations to enter the reasoning process quickly~\cite{llm_agents_6g}. In such scenarios, the reasoning process must account for wireless network connectivity because inter-agent information exchange is carried over wireless links. Poor connectivity can delay or interrupt the information required for subsequent reasoning steps, thus degrading the performance of task completion~\cite{integrated_ai_comm}.

However, conventional agent systems and wireless communication networks are generally designed separately. On the one hand, an LLM-based MAS organizes collaboration through an interaction structure that determines which agents exchange information during reasoning. At the initial stage, this structure was prescribed in advance through fixed roles and predefined interaction patterns~\cite{autogen,jiang2024}. As collaborative tasks become more diverse, such fixed structures cannot adapt to the current reasoning demand. Recent work has therefore made the interaction structure adaptive. For instance, AgentPrune removes exchanges with low task contribution~\cite{agentprune}, while G-Designer generates task-dependent interaction topologies from the current collaboration context~\cite{gdesigner}. Nonetheless, this formation process still largely ignores the network architecture and connectivity through which the selected interactions must be realized.

On the other hand, conventional wireless network design typically assumes a given network topology and communication demand, and then allocates wireless resources to support the corresponding transmissions. Conventionally, wireless design focused on reliable bit delivery under channel and resource constraints. Recently, task-oriented and pragmatic communication introduced task utility into transmission design~\cite{hu2026pragcomm}, but these approaches mainly optimize how a given interaction is delivered at the link level. They do not determine how inter-agent collaboration should be formed in an MAS. More recently, distributed LLM inference has been designed coupling with wireless transmission with edge computation~\cite{zhang2025dist_inference}, while agentic communication has begun to relate communication decisions to the reasoning process~\cite{reasoning_native_6g}. Recent studies have also considered multi-agent coordination over 6G networks~\cite{jiang2024,llm_agents_6g}. Nevertheless, how agent interactions should be formed according to both reasoning requirements and network conditions remains not well understood.

Motivated by this gap, this article studies joint agent and network design for LLM-based MASs over wireless networks. We identify six technical challenges, including metric misalignment and
message redundancy, state inconsistency and topology mismatch, as well as
resource limitation and trust discontinuity. We then develop various joint design approaches that use task utility to connect agent-side decisions with network-side communication and resource allocation. An illustrative vehicle-to-everything (V2X) case study compares the joint design with two conventional separate designs, where the agent-only design adapts the agent-side information decision with fixed network resources, while the wireless-only design allocates network resources for a fixed agent-side decision. The results show that the joint design improves task completion under communication and edge-resource constraints and outperforms both separate designs.

\begin{table*}[t]
\centering
\caption{Joint design space for network-constrained LLM-based MASs.}
\label{tab:joint_design_space}
\renewcommand{\arraystretch}{1.20}
\setlength{\tabcolsep}{5pt}
\footnotesize

\begin{tabularx}{\textwidth}{
>{\raggedright\arraybackslash}p{0.17\textwidth}
>{\raggedright\arraybackslash}p{0.165\textwidth}
>{\raggedright\arraybackslash}p{0.185\textwidth}
>{\raggedright\arraybackslash}p{0.195\textwidth}
>{\raggedright\arraybackslash}X}
\toprule

\textbf{Challenge}
&
\textbf{Agent Side}
&
\textbf{Network/Edge Side}
&
\textbf{Coupling}
&
\textbf{Joint Decision}
\\
\midrule

Metric misalignment
&
Interaction priority~\cite{credit_assignment}
&
Resource allocation
&
Interaction credit $\leftrightarrow$ service feasibility
&
Interaction--resource scheduling
\\

Message redundancy and state inconsistency
&
Message selection/compression~\cite{agentprune}
&
Delivery fidelity/timeliness~\cite{hu2026pragcomm,reasoning_native_6g}
&
Receiver-context-conditioned decision distortion
&
Message selection and adaptive delivery
\\

Topology mismatch
&
Logical edge/endpoint selection~\cite{gdesigner,agentprune, darwin_godel}
&
Association/path selection~\cite{integrated_ai_comm,reasoning_native_6g}
&
Interaction credit $\leftrightarrow$ support cost
&
Joint edge--network mapping
\\

Resource limitation
&
Reasoning budget/workload
&
Communication/edge-resource allocation
~\cite{zhang2025dist_inference,llm_agents_6g}
&
Workload footprint $\leftrightarrow$ resource feedback
&
Workload--resource allocation
\\

Trust discontinuity
&
Information-flow control~\cite{fides}
&
Provenance verification/admission
&
Provenance $\leftrightarrow$ trust requirement
&
Provenance-aware information flow
\\

\bottomrule
\end{tabularx}
\end{table*}

\section{Network-Constrained LLM-Based MASs}
\label{sec:system}

This section establishes the system foundation for LLM-based MASs over wireless networks. We first describe an individual LLM agent and then extend to multi-agent collaboration over wireless network edge\footnote{The discussion also applies to agentic systems built on other multimodal foundation models, including multimodal large language models (MLLMs), vision-language models (VLMs), as well as vision-language-action (VLA) models. A detailed treatment of multimodal perception and action is beyond the scope of this article.}. We then discuss representative
applications and the resulting technical challenges.
\subsection{System Basics}
\label{subsec:foundations}

An LLM-based agent is an autonomous decision entity that uses an LLM as its
reasoning core, with memory and external tools supporting its operation, as
illustrated in the left panel of Fig.~\ref{fig:system}. At each reasoning step,
the agent forms a \textit{context}, defined as the information currently
available to the LLM from its input and retained memory. The LLM processes this
context to determine the next action, and the agent then updates the context
based on newly obtained information. This process continues as the task
progresses.

An LLM-based MAS consists of multiple LLM agents that cooperate on a shared
task. One important issue faced by such
systems is \textit{reasoning dependency}, i.e., one agent generally needs to acquire the information or a processing
result from another before the corresponding reasoning
step can continue. Here, the message exchanged to
satisfy such a dependency is referred to as an \textit{interaction}. When the agents are hosted on different network
nodes, these messages are exchanged over wireless communication links.
Wireless connectivity can therefore affect when the required information
reaches the receiving agent and, consequently, whether the dependent reasoning
step can proceed. Since wireless links may vary during task execution, changes
in message delivery can further affect the subsequent multi-agent
collaboration.

LLM agents can be hosted at different locations across the network (e.g., at end devices, edge nodes, and cloud servers). Agents
hosted on end devices remain close to local observations and physical actions,
but have limited computing capacity. Agents hosted at nearby edge nodes can
access additional edge computing resources while remaining close to the devices
they support. Remote cloud servers can accommodate larger models or longer
context, but the longer communication path makes cloud execution less suitable
for time-critical reasoning. Agents hosted across these tiers can also
collaborate during task execution, allowing latency-sensitive reasoning to
remain close to the information source while more resource-demanding reasoning
is supported by edge or cloud agents. Such cross-tier collaboration therefore
creates a tradeoff between communication distance and available execution
resources, which affects whether a reasoning dependency can be completed
within the task deadline.

The right panel of Fig.~\ref{fig:system} illustrates how multiple agents
collaborate during task execution. We use the term \textit{collaboration
workflow} to denote the runtime process that converts a shared task into
reasoning dependencies and maps these dependencies to multi-agent interactions.
Newly obtained information may change the unresolved dependencies during
execution, causing the interaction structure to be updated. The collaboration
workflow therefore determines how inter-agent collaboration evolves as the
task progresses.

Table~\ref{tab:joint_design_space} summarizes the coupling between this
workflow and network operation. A reasoning dependency does not necessarily
correspond to a unique interaction because different agents may provide
information that satisfies the same task requirement. The MASs determine which
interaction is useful for the current reasoning process, while the network
determines whether that interaction can be supported under the available
connectivity and resources. Task execution is therefore constrained by both
the multi-agent interaction workflow and the underlying network, which raises
the joint agent and network design challenges, as will be discussed in Section~\ref{subsec:challenges}.

\begin{figure*}[t]
    \centering
    \includegraphics[width=0.98\textwidth]{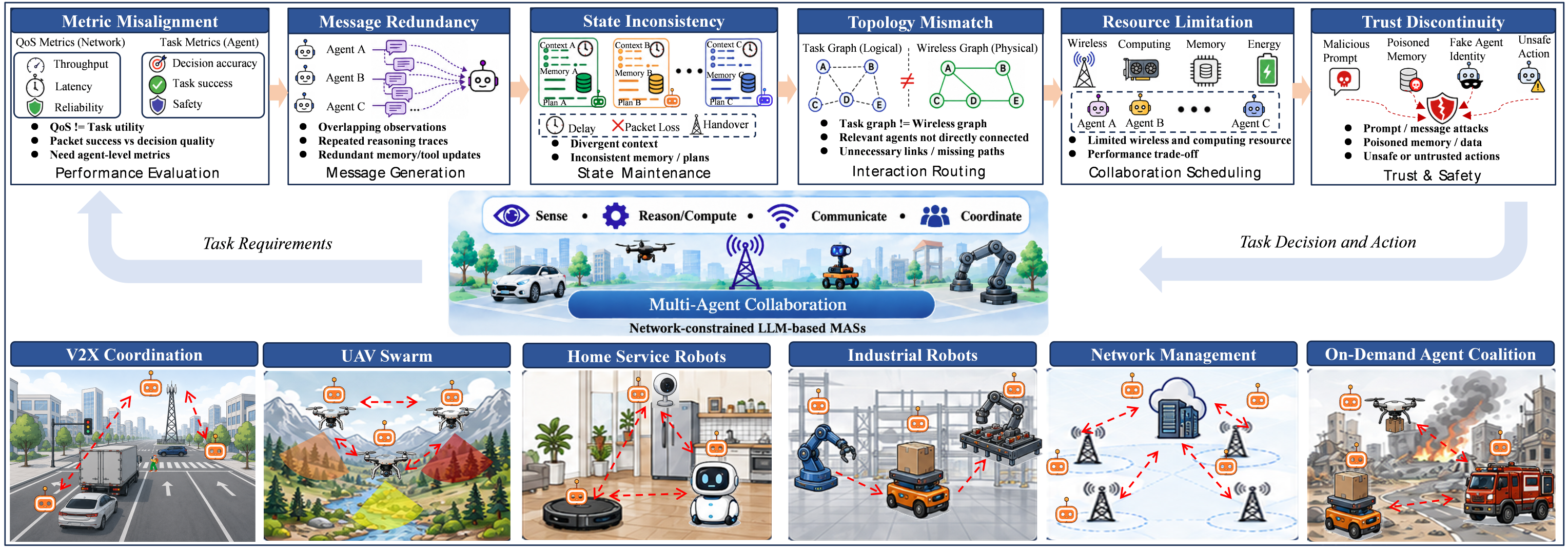}
    \caption{Key challenges and representative applications of LLM-based MASs over wireless networks.}
    \label{fig:challenges-applications}
\end{figure*}

\subsection{Representative Applications}
\label{subsec:applications}

Fig.~\ref{fig:challenges-applications} summarizes representative applications of LLM-based MASs over wireless networks.

\textbf{Vehicle-to-Everything (V2X) Coordination:}
In cooperative perception, an ego vehicle may require observations from nearby nodes to identify an occluded pedestrian or vehicle. An LLM agent hosted on a nearby vehicle or roadside unit (RSU) can provide the missing information, while the wireless network determines whether the message can reach the ego vehicle within the required decision time.

\textbf{Uncrewed Aerial Vehicle (UAV) Swarm:}
In a search or monitoring mission, UAV agents can divide the sensing area and exchange local observations during execution. When a UAV completes its assigned region or loses a reliable link, the remaining agents may need to redistribute the sensing task. The resulting collaboration therefore depends on both the evolving mission and the communication links among the UAVs.

\textbf{Networked Robotics:}
In networked robotic systems, agents may be hosted on manipulators, mobile robots, or local controllers that perform different parts of a shared task. A robot may require an observation or execution result from another robot before its next action can proceed. LLM-based MASs can adapt the subsequent task assignment according to these intermediate results, while wireless connectivity determines whether the required information is available in time.

\textbf{Network Management:}
In this scenario, the wireless network serves as both the communication infrastructure and the object being controlled. Distributed agents can diagnose a network problem and modify its configuration. The resulting network conditions then affect the communication available for subsequent agent collaboration.

\textbf{On-Demand Agent Coalition:}
In emerging tasks such as disaster response or emergency search, the participating agents may not be known in advance. An LLM-based MAS can identify available agents according to the current task and form a temporary coalition using agents hosted on mobile or infrastructure nodes. The wireless network determines which of these candidate agents can participate under the current connectivity and resource conditions.

\subsection{Technical Challenges}
\label{subsec:challenges}

Network-constrained LLM-based MASs face various technical challenges when
multi-agent reasoning is executed over wireless networks. As summarized in
Fig.~\ref{fig:challenges-applications}, these challenges arise from task
evaluation and information exchange, as well as from the network realization
and trusted execution of agent interactions.\footnote{The six challenges span
task evaluation and information transfer, interaction realization and
resource-constrained execution, as well as trusted information use. Message
redundancy and state inconsistency are addressed together because both concern
whether task-relevant information reaching the receiver remains useful for its
subsequent reasoning.}

\textit{1) Metric Misalignment:}
Wireless systems typically characterize transmission performance using
quality-of-service (QoS) metrics such as throughput and delay. These metrics
describe how effectively information is delivered, but not how much a
particular interaction contributes to the multi-agent task. This contribution
is difficult to measure, as its effect may become observable only after
subsequent reasoning has unfolded, whereas wireless scheduling must prioritize
transmissions before the task outcome is known. The network therefore lacks a
directly measurable interaction-level metric that connects immediate
transmission decisions to downstream task utility.

\textit{2) Message Redundancy:}
Message redundancy occurs when agents exchange observations or reasoning
results that add little new information to the receiver~\cite{agentprune}.
Over wireless links, such redundant message transmission consumes
communication resources and may delay more useful interactions. Reducing such
redundancy is essential but difficult because content similarity does not
determine a message's contribution to the receiver's next decision. A message
similar to the existing context may provide a missing fact, while a different
message may have little effect on the current reasoning. Its usefulness must
therefore be evaluated with respect to the receiver's current context and
subsequent decision.

\textit{3) State Inconsistency:}
The \textit{state} of an agent refers to the context used for its next
reasoning step. Here, \textit{context} denotes the information available to the
LLM for reasoning, while \textit{state} emphasizes the information whose
consistency affects coordinated execution. Different agents may retain
inconsistent states when a shared update is delayed or missed during wireless
transmission. Nevertheless, reliable delivery alone does not remove this
problem because an update may arrive after the receiving agent has already
reasoned from an outdated state. Maintaining the freshness of message updates
is therefore important but difficult, as the required freshness depends on how
an update affects the subsequent reasoning step. Packet delivery status alone
does not reveal this requirement.

\textit{4) Topology Mismatch:}
An MAS forms logical interactions according to reasoning dependencies, while
the wireless network provides physical connectivity between the nodes hosting
the agents. A topology mismatch occurs when a task-relevant agent is difficult
to reach through the available links, whereas a well-connected agent may
provide little useful information~\cite{llm_agents_6g,gdesigner}. Therefore,
interaction selection is important but challenging, as it depends on both task
relevance and physical reachability. Neither the agent interaction graph nor
the wireless connectivity alone determines which interaction should be
realized.

\textit{5) Resource Limitation:}
Distributed LLM agents rely on local sensing capabilities, while communication
and computation resources at edge nodes are limited
~\cite{llm_agents_6g,zhang2025dist_inference}. The resulting workload is not
fully known before execution because received context can change the amount of
subsequent reasoning. Intermediate reasoning results may also trigger new
inter-agent exchanges. Consequently, communication and execution demands can
evolve during the collaboration itself. Allocating communication and
computation resources is therefore essential but difficult because the demand
being served is partly generated by the same reasoning process whose execution
depends on those resources.

\textit{6) Trust Discontinuity:}
Wireless security mechanisms can authenticate a sender and verify the integrity
of a delivered message, but these guarantees do not determine how the received
information should be allowed to influence subsequent agent reasoning and
actions. The corresponding design issue is therefore to preserve
network-verified provenance and use it to constrain downstream information
use. This is important because an authenticated message may still contain a
prompt injection or other malicious instruction that alters the receiver's
behavior~\cite{prompt_infection}. It is also difficult because provenance
established during delivery may no longer remain explicit after the
information is incorporated into the agent's context or memory. Network-side
verification cannot determine the later semantic influence of the message,
while agent-side control may lack the provenance needed to enforce
source-dependent policies. Secure multi-agent reasoning therefore requires
network-established provenance to remain available as information propagates
through the agent's reasoning and execution process~\cite{fides}.

\section{Joint Agent and Network Design}
\label{sec:joint_design}

To resolve the above challenges, we pursue joint agent and network design
based on task requirements and network conditions. Task utility serves as the primary performance objective, while trust
requirements impose constraints on how information can influence agent
operations. The other designs identify the coupled agent-side and
network-side decisions under these objectives and constraints.

\subsection{Joint Interaction Scheduling and Resource Allocation}
\label{subsec:credit}

Metric misalignment requires interaction scheduling and network resource
allocation to be considered jointly. An MAS may select an interaction with a
high contribution to task completion even when the communication link
supporting that interaction cannot deliver its message before the task
deadline. Conversely, a network-side scheduler that considers only queue and
channel conditions may allocate more resources to a link carrying an
interaction with little contribution to task completion. Hence, interaction
activation and resource allocation should account for both task contribution
and link feasibility.

Credit assignment provides the task-side metric needed for this decision. In
MASs, \textit{credit} denotes the contribution of an agent or action to a
shared task outcome~\cite{credit_assignment}. We extend this notion to an
inter-agent exchange and define its \textit{interaction credit} as the
estimated marginal contribution of that exchange to task utility under a
given delivery condition. Offline evaluations can approximate this
contribution by comparing task outcomes when the interaction is available
with those obtained when it is withheld or delayed under otherwise matched
settings. A runtime predictor can then estimate the interaction credit from
the current task context and expected delivery condition. The MASs use the
predicted credit to prioritize candidate interactions, while the network
provides the corresponding link feasibility or service cost. These quantities
jointly determine which interactions are activated and how resources are
allocated. A modified max-weight policy provides one possible implementation
by incorporating interaction credit into the conventional service weight.

\subsection{Joint Message Selection and Transmission}
\label{subsec:info_state}

Message redundancy and state inconsistency require the system to determine
what information should be transmitted and how it should be delivered.
Agent-side message selection alone may retain information that is useful in
isolation but redundant to the receiver's current context or ineffective when
delivered too late. Network-side transmission, in contrast, treats the
selected content as given and does not know which part will affect the
receiver's next reasoning step. A joint design therefore requires a common
receiver-side criterion for message selection and transmission.

Communication-efficient methods provide the agent-side mechanism for selecting
message content~\cite{agentprune}, while task-oriented joint source--channel
coding (JSCC) provides a network-side mechanism for adapting transmission to
an end-task loss~\cite{hu2026pragcomm,reasoning_native_6g}. These mechanisms
can be coupled through a \textit{receiver-context-conditioned decision
distortion}, which measures how removing, delaying, or distorting information
changes the receiver's subsequent decision. The MASs use this metric to retain
or compress information according to its decision impact. The transmission
scheme uses the same metric to determine the required delivery fidelity. For
a time-sensitive state update, delayed delivery produces a larger penalty when
the update no longer supports the receiver's next reasoning step.

\subsection{Joint Agent and Network Topology Design}
\label{subsec:topology}

Topology mismatch requires the logical interaction graph and its network
realization to be configured together. Adaptive MASs can construct or prune
interaction graphs according to reasoning dependencies, with the interaction
credit introduced above representing the task value of each candidate edge
~\cite{gdesigner,agentprune}. Wireless association and routing instead
determine how the nodes hosting the selected agents can be connected under
current channel and resource conditions. If these decisions are separated, a
high-value logical edge may be difficult to realize, while a network-efficient
connection may correspond to an interaction with little task value. Task
relevance and network feasibility should therefore enter the same topology
decision.

For each unresolved reasoning dependency, the MASs identify candidate agents
and assigns an interaction credit to the corresponding logical edges. The
network then evaluates feasible associations or communication paths for each
candidate and computes a \textit{support cost}, defined as the communication
resources required to satisfy its delivery constraints. The logical edge and
its network realization are subsequently selected according to interaction
credit and support cost under the available connectivity and resources.
Because both choices are discrete, the resulting edge--network mapping forms
a combinatorial optimization problem. Existing task-dependent graph
construction and wireless association or routing provide the two building
blocks for this joint topology design.

\subsection{Joint Workload Control and Resource Allocation}
\label{subsec:workload}

Resource limitation couples the workload generated by the MASs with the
resources available to execute it. An agent-side controller may select a more
demanding reasoning configuration for greater expected task gain even when the
resulting workload cannot be completed within the available resources and task
deadline. Conversely, conventional resource allocation treats the submitted
workload as fixed or uncontrollable and only determines how the available
resources should serve it. Workload control and resource allocation should
therefore be determined together according to expected task gain and execution
feasibility.

Here, \textit{agentic workload} denotes the communication and execution demand
induced by a reasoning configuration. Reasoning-budget control provides one
agent-side mechanism for changing this workload, where the
\textit{reasoning budget} bounds the extent of reasoning performed by the
agents. Distributed inference and computation offloading provide network-side
mechanisms for allocating communication and edge resources
~\cite{zhang2025dist_inference,llm_agents_6g}. A workload profiler can connect
the two by predicting the resource footprint induced by a reasoning
configuration without executing the complete reasoning process. The reasoning
configuration and resource allocation can then be jointly optimized to
maximize expected task utility under the task deadline and the available
communication and computing resources. For larger design spaces, marginal
resource prices derived from active constraints can provide compact feedback
for adapting the reasoning budget as network and edge conditions change.

\subsection{Joint Provenance Verification and Information-Flow Control}
\label{subsec:security}

Trust discontinuity requires network-side provenance verification to remain
connected with agent-side control over how received information is used.
Network authentication and integrity verification establish the source and
integrity of a delivered message, while information-flow control (IFC)
restricts how that information can propagate through reasoning and execution
~\cite{fides}. If these mechanisms operate separately, an authenticated
message may still contain malicious instructions~\cite{prompt_infection}, and
the agent may later lack the verified provenance needed to enforce
source-dependent policies. Joint design therefore needs to preserve verified
provenance after delivery and enforce it when the information influences agent
operations.

One implementation is to bind the verified source and integrity attributes of
an incoming message to its content as a \textit{provenance label}. Agent-side
IFC propagates this label when the information enters context or memory and
when it contributes to derived results. Before a protected operation, the
agent checks whether the provenance associated with the influencing
information satisfies the corresponding policy. The coupling can also operate
in the reverse direction. If a reasoning dependency requires information from
a specified trust class, this requirement can constrain source selection or
message admission. Network verification and agent-side IFC therefore connect
the origin of received information with the permissions governing its
downstream influence.

\begin{figure}[t]
    \centering
    \includegraphics[width=0.99\linewidth]{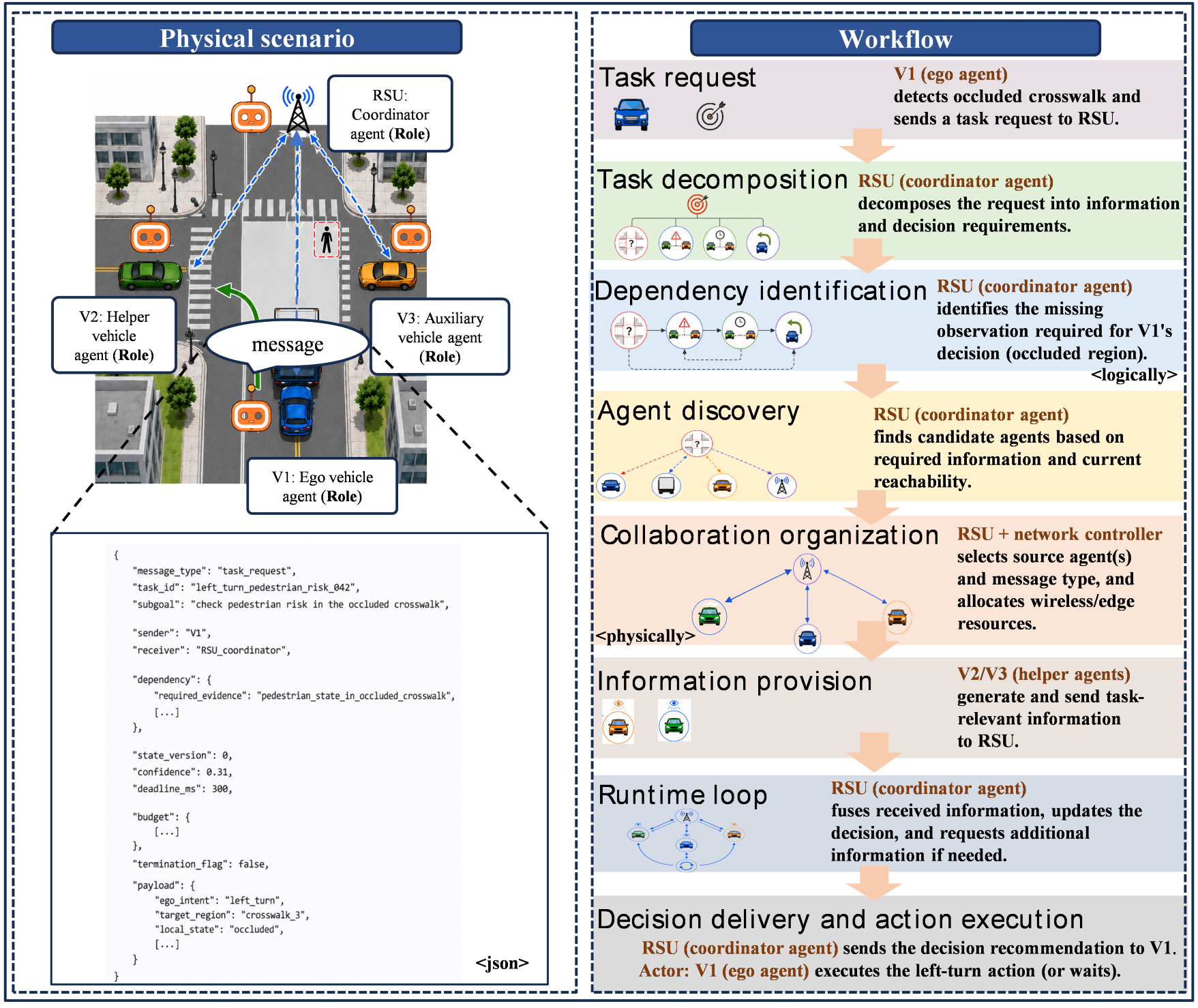}
    \caption{Illustration of case scenario and collaboration workflow.}
    \label{fig:case-v2x}
\end{figure}

\section{Illustrative Case Study: V2X Multi-Agent Collaboration}
\label{sec:case}

To illustrate joint agent and network design, we consider a concrete V2X task
in an unprotected left-turn scenario with an occluded pedestrian, as shown in
Fig.~\ref{fig:case-v2x}. The ego vehicle $V_1$ must decide whether it is safe
to enter the conflict zone, while a nearby truck blocks its view of the
pedestrian. Vehicle $V_2$ has a complementary view of the occluded region,
whereas $V_3$ provides additional traffic information~\cite{v2xsim}. Each
vehicle hosts an LLM-based agent, while the RSU serves as a base station
(BS)-like wireless access point and hosts the coordinator agent. All vehicles
communicate with the RSU through vehicle-to-infrastructure links; no direct
vehicle-to-vehicle (V2V) or device-to-device (D2D) link is considered. This
physical topology remains fixed for all evaluated designs, so the case study
focuses on interaction, message, and resource decisions rather than topology
optimization. Vehicle-side execution resources are fixed, while the uplink
bandwidth and the RSU computation and memory resources constrain task
execution.

Fig.~\ref{fig:case-v2x} further specifies which actor performs each stage of
the collaboration. First, the LLM agent on $V_1$ detects that its local
observation is insufficient and sends a task request to the RSU. The RSU agent
then decomposes the request, identifies the missing observation required for
the turning decision, and discovers candidate agents on nearby vehicles.
Next, the RSU coordinator and network controller determine the information
source and message configuration according to the current design and resource
conditions. The selected helper agent uses its local observation and LLM to
generate task-relevant information and sends the resulting message to the RSU.
Finally, the RSU agent incorporates the received information into the current
task context. If the information remains insufficient, it initiates another
interaction; otherwise, it sends the safety recommendation to $V_1$, which
executes or postpones the left-turn action.

In simulations, an interaction is specified by the selected source agent and
the amount of task-relevant information included in its message. The joint
design additionally determines the wireless resources allocated to deliver the
selected message. We use successful task completion as the task utility and
report its average completion rate. A task is completed only when the selected
information is sufficient for the safety decision and can be delivered and
processed within the prescribed end-to-end deadline, which captures the
latency constraint of the V2X task. The source and message decisions, as well
as wireless resource allocation, are subject to the available uplink bandwidth
and the RSU computation and memory constraints. We compare three designs under
the same physical topology. The \textit{agent-only} design adapts the source
agent and message content with fixed wireless resource allocation. The
\textit{wireless-only} design fixes the source agent and message content while
adapting the wireless resource allocation. The \textit{joint} design adapts
the source and message together with the wireless resource allocation while
accounting for the RSU execution constraints.

\begin{figure}[t]
    \centering
    \includegraphics[width=0.98\linewidth]{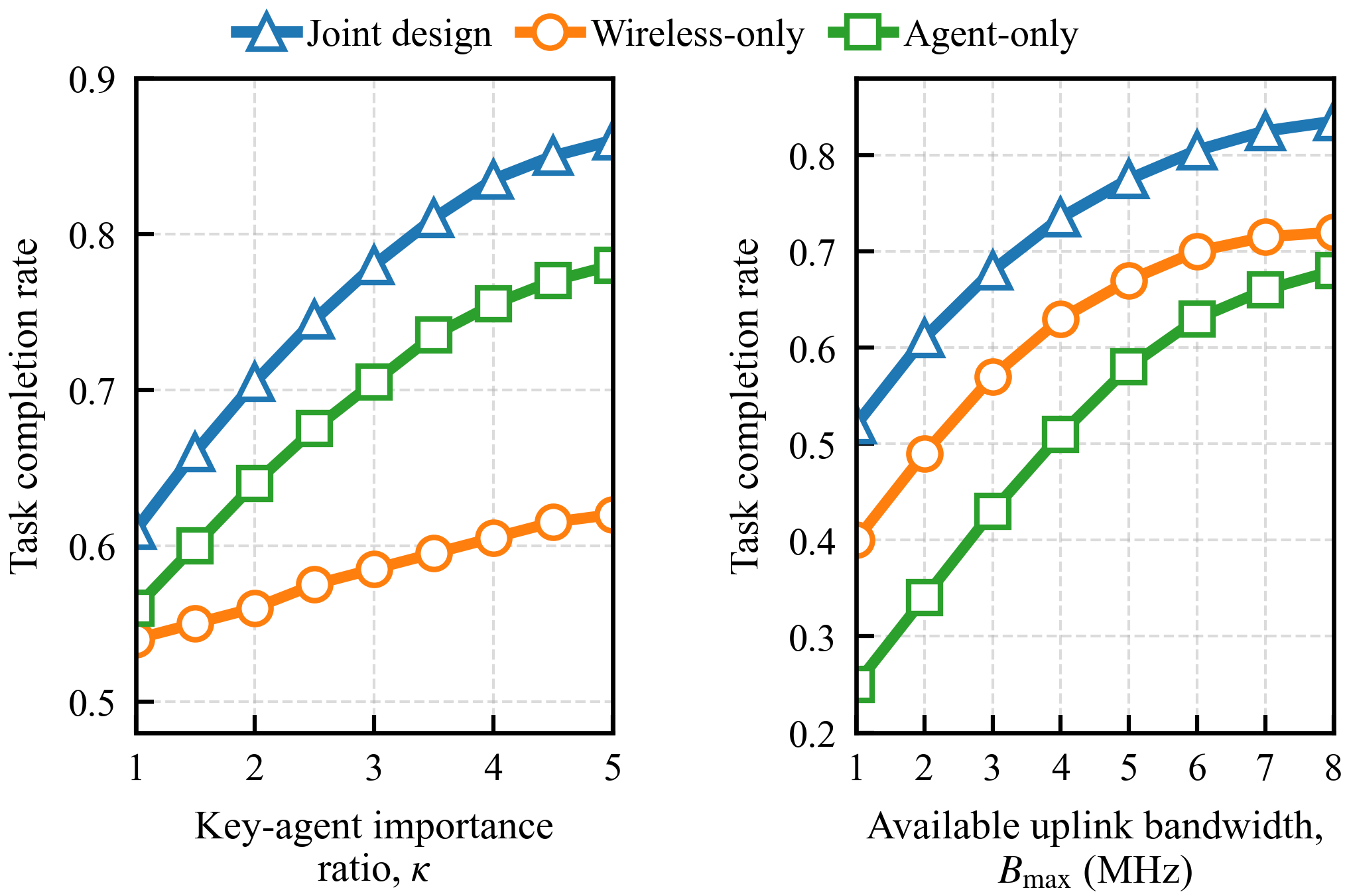}
    \caption{Task completion rate under two bottleneck regimes. Under the information-limited regime (left), $\kappa$ measures the relative decision relevance of the key agent; under the bandwidth-limited regime (right), $B_{\max}$ denotes the available uplink bandwidth.}
    \label{fig:task-completion-two-regimes}
\end{figure}

Fig.~\ref{fig:task-completion-two-regimes} examines two communication regimes.
In the information-limited setting, we define the \textit{key agent} as the
candidate agent whose observation is most relevant to the current safety
decision. The key-agent importance ratio $\kappa$ characterizes the
contribution of this agent relative to the other candidate agents. As
$\kappa$ increases, selecting the key agent becomes more important for
successful task completion. The wireless-only design cannot exploit this
difference because its information source is fixed. The agent-only and joint
designs can select the more useful source, while the joint design also accounts
for the wireless feasibility of the resulting message. In the
bandwidth-limited setting, the useful source is fixed and the available uplink
bandwidth $B_{\max}$ is varied. The bottleneck therefore shifts to message
delivery. Increasing $B_{\max}$ improves the support available to the selected
interaction, while the joint design can further adapt the transmitted
information amount to the available bandwidth.

\begin{figure}[t]
    \centering
\includegraphics[width=0.98\linewidth]{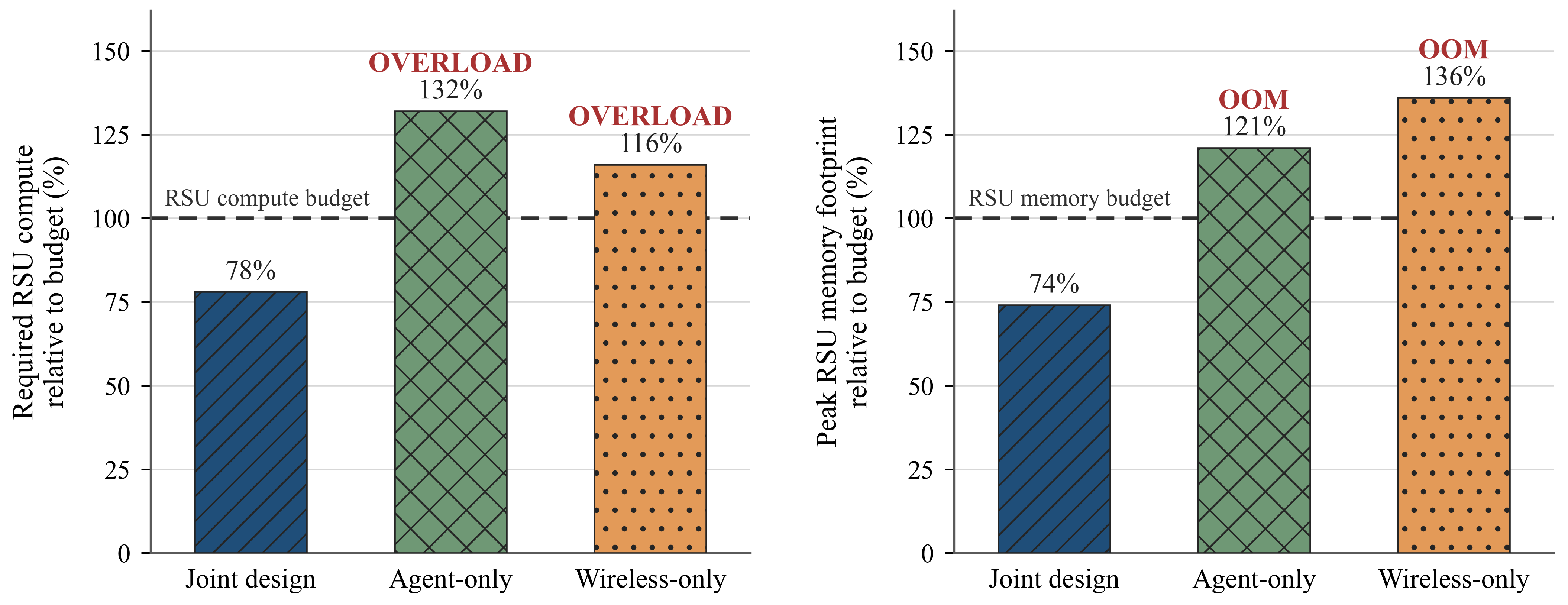}
    \caption{Compute and memory bottlenecks under different agent--network designs (OOM is out of memory).}
    \label{fig:failure-modes}
\end{figure}

Fig.~\ref{fig:failure-modes} examines the execution constraints at the RSU.
The joint design achieves a higher task completion rate than the agent-only and
wireless-only designs under both compute- and memory-limited settings. The RSU
computation and memory demands are interaction-dependent: changing the source
or message amount changes the information processed at the RSU and hence the
resulting execution workload. The processing budget specifies the computation
available before the task deadline, while the memory budget limits the amount
of execution memory available at the RSU. The agent-only design may select an
informative interaction whose resulting workload exceeds these constraints
because execution feasibility does not enter its interaction decision. The
wireless-only design can improve delivery of the predefined interaction but
cannot change the computation or memory demand created after reception. The
joint design can instead reduce the message amount or select another source
when resources become limited, thereby adapting the generated workload to the
available RSU resources.

\section{Conclusion}
\label{sec:conclusion}

This article examined network-constrained LLM-based MASs from a joint agent
and network design perspective. We showed that agent-side decisions and
network operation should be jointly designed across interaction scheduling,
information transfer, topology configuration, workload execution, and
cross-layer trust. The V2X case study illustrated how jointly adapting
interaction and resource decisions improves task completion under
communication and edge-resource constraints. These results motivate future
wireless systems to coordinate agent collaboration with network
connectivity and edge resources according to task requirements.

\end{document}